\documentclass[twoside]{article}
\usepackage{amssymb}
\usepackage{qic}
\usepackage{graphicx} 
\usepackage{subfigure}
\usepackage{dcolumn}  
\usepackage{bm}       
\usepackage{amsmath}
\usepackage{graphics}
\usepackage{amscd}
\usepackage{afterpage}
\usepackage{float,times}
\usepackage{epsfig}
\usepackage{theorem}
\usepackage{wrapfig}
\usepackage{picinpar}

\DeclareSymbolFont{AMSb}{U}{msb}{m}{n}
\DeclareMathSymbol{\N}{\mathbin}{AMSb}{"4E}
\DeclareMathSymbol{\Z}{\mathbin}{AMSb}{"5A}
\DeclareMathSymbol{\R}{\mathbin}{AMSb}{"52}
\DeclareMathSymbol{\Q}{\mathbin}{AMSb}{"51}
\DeclareMathSymbol{\I}{\mathbin}{AMSb}{"49}
\DeclareMathSymbol{\C}{\mathbin}{AMSb}{"43}

\def\R{\mathbb{R}}
\def\N{\mathbb{N}}
\def\C{\mathbb{C}}
\def\Z{\mathbb{Z}}

\begin{document}
\setlength{\textheight}{8.0truein}    

\runninghead{Gate-level simulation of logical state preparation}
            {A.M. Stephens, S.J. Devitt, A.G. Fowler, J.C. Ang, and L.C.L. Hollenberg}

\normalsize\textlineskip
\thispagestyle{empty}
\setcounter{page}{1}


\vspace*{0.88truein}

\alphfootnote

\fpage{1}

\centerline{\bf
GATE-LEVEL SIMULATION OF LOGICAL STATE PREPARATION}

\vspace*{0.37truein}
\centerline{
A.M. Stephens$^{1}$, S.J. Devitt$^{1,2}$, A.G. Fowler$^{3}$, J.C. Ang$^{1}$, and L.C.L. Hollenberg$^{1}$}

\vspace*{0.119truein}
\centerline{$^{1}$\it Centre for Quantum Computer Technology, 
School of Physics, The University of Melbourne}

\baselineskip=10pt
\centerline{\it Melbourne, Victoria 3010,
Australia}

\vspace*{0.119truein}
\centerline{$^{2}$\it Centre for Quantum Computation}
\centerline{\it Department of Applied Mathematics and Theoretical Physics,
University of Cambridge
}

\baselineskip=10pt
\centerline{\it Cambridge CB3 0WA,
United Kingdom
}

\vspace*{0.119truein}
\centerline{$^{3}$\it Institute for Quantum Computing, University of Waterloo
}

\baselineskip=10pt
\centerline{\it Waterloo, Ontario N2L 3G1,
Canada
}

\vspace*{10pt}

\vspace*{0.21truein}

\abstracts{
Quantum error correction and fault-tolerant quantum computation are two fundamental concepts which make quantum computing feasible. While providing a theoretical means with which to ensure the arbitrary accuracy of any quantum circuit, fault-tolerant error correction is predicated upon the robust preparation of logical states. An optimal direct circuit and a more complex fault-tolerant circuit for the preparation of the [[7,1,3]] Steane logical-zero are simulated in the presence of discrete quantum errors to quantify the regime within which fault-tolerant preparation of logical states is preferred.
}{}{}

\vspace*{10pt}

\vspace*{3pt}

\vspace*{1pt}\textlineskip   

\section{Introduction}

Faced with the difficulty inherent to the simulation of quantum mechanical systems using classical computers, it was Feynman who first suggested using quantum mechanics itself as a basis for computation \cite{Feynman1}.  While this idea was formalised by Deutsch, who conceived the universal quantum computer as one which would facilitate the efficient simulation of an arbitrary physical system \cite{Deutsch1}, algorithms developed by Simon \cite{Simon1} and Shor \cite{Shor1} more explicitly demonstrated the potential of the quantum computer, precipitating a number of initial proposals aimed at the physical construction of a working device \cite{Cirac1, Cory1, Gershenfeld1, Barenco1, Kane1, Loss1}. 
\\
\\
Despite this progress, one significant problem which remains an impediment to the realisation of any form of quantum computing is that of quantum errors, which can result from both environmental decoherence and systematic imprecision. Quantum error correction (QEC), devised independently by Shor \cite{Shor2} and by Steane \cite{Steane1} and later generalised by Calderbank and Shor \cite{Calderbank1} and Steane \cite{Steane2}, initially established a means of negating errors afflicting a static qubit. Subsequently, the theory of fault-tolerant quantum computation demonstrated that, by controlling the propagation of errors, QEC can be undertaken dynamically and in spite of faulty circuit components or externally induced errors \cite{DiVincenzo2, Gottesman1, Shor3}. In promising increasingly accurate operation with each level of concatenation, provided that the underlying physical error rate is below some threshold, $p_{th}$, fault-tolerant quantum computation quickly became the theoretical basis of conventional quantum computing.
\\
\\
Given its importance, a large body of research has endeavored to quantify the threshold error rate for fault-tolerant quantum computation. While this work has comprised some analytic results \cite{Preskill2, Gottesman2, Gottesman1, Aharonov1}, classical simulation has shown also to be an effective method of threshold estimation. One approach is to simulate the propagation of errors throughout a circuit, rather than the evolution of the entire system state \cite{Steane4}. Several papers have adopted this method for the purposes of threshold estimation \cite{Zalka1, Salas1} and in the evaluation of various ancilla preparation schemes \cite{Salas2, Reichardt1}. Correction circuits based upon the stabilizer formalism \cite{Gottesman1} are particularly conducive to simulation using a package such as CHP (CNOT-Hadamard-Phase) \cite{Aaronson1}, though such a method is not applicable to non-stabilizer codes or true quantum algorithms as the simulation of any stabilizer circuit can be performed efficiently on a classical computer \cite{Gottesman3}. Other authors have demonstrated the effectiveness of a state vector approach in simulating quantum circuits \cite{Fowler1, Devitt3}. However, as errors are modeled by the stochastic application of appropriate gates, the time required to generate statistically significant results becomes prohibitive in the low $p$ regime. As an alternative, the density matrix formalism explicitly includes the stochastic averaging of quantum errors. The limiting factor then becomes the amount of physical memory available to store and manipulate the system density matrix. This approach of gate-level simulation forms the foundation for this research, and is also the basis for other general purpose quantum computer simulators including QuIDD (Quantum Information Decision Diagrams) \cite{Viamontes1, Viamontes2, Viamontes3}, and is applicable to the simulation of algorithms  such as Shor's \cite{Ang1}.
\\
\\
Despite the theoretical potential of fault-tolerance to reduce arbitrarily the effect of induced errors, fault-tolerant error correction is predicated upon the encoding of physical qubits to form logical states. Unlike logical gate operation, error correction cannot be undertaken until a complete logical state is encoded. That is, correction cannot be performed after every time step of a preparation circuit. Concatenation further strengthens the requirement of robust logical preparation, as the number of required logical states grows exponentially with each level of recursive encoding. In addition, ancilla networks have been proposed which require logical states as the central resource \cite{Steane5}.  While an encoding circuit can itself adhere to the conditions of fault-tolerance, insisting upon fault-tolerant operation will necessarily increase both the number of qubits and the number of gate operations required. This increase in circuit area corresponds directly to an increased logical region within which an error can occur. In contrast, encoding undertaken directly, while not being strictly fault-tolerant, will be vastly simpler in its implementation and will, therefore, be more rapid and less susceptible to the initial occurrence of quantum errors. In practice, and prior to any concatenation, this balance between circuit complexity and the degree of effective error protection implies a crossover error rate, $p_{cr}$, at which the direct and fault-tolerant implementations of an arbitrary circuit become effectively equivalent. This paper involves the gate-level simulation of direct and fault-tolerant circuits for the preparation of the [[7,1,3]] Steane logical-zero to provide a quantitative determination of the crossover error rate, below which fault-tolerant quantum computation becomes the preferred method of ensuring accurate state preparation.
\\
\\
The structure of this paper is as follows. Sections \ref{QEC} and \ref{LSP} comprise a brief overview of the relevant principles underlying both the Steane code and the circuits for logical state preparation. Section \ref{SM} details the error model and other technical issues relating to simulation. Results are tabled in Section \ref{A}, whereupon the crossover physical error rate is presented.

\section{The Steane code}
\label{QEC}

Though the five qubit code \cite{Bennett1, Knill1} represents the most compact QEC protocol, the seven qubit Steane code  \cite{Steane1, Steane2} can be considered a more practical code as universal fault-tolerant computation can be performed more simply \cite{Gottesman1}. The Steane code is a sufficient protocol to successfully protect against an arbitrary single qubit error, and even multiple errors with non-zero probability of success \cite{Fern1}. The encoding procedure for the Steane code consists of the transformation $\alpha\vert0\rangle+\beta\vert1\rangle\rightarrow\alpha\vert0_{L}\rangle+\beta\vert1_{L}\rangle$, where the logical codewords, $\vert0_{L}\rangle$ and $\vert1_{L}\rangle$, each comprise a weighted eight-state superposition, and are given by
\begin{equation}
{\setlength\arraycolsep{1pt}
\begin{array}{rl}
\vert0_{L}\rangle=\frac{1}{\sqrt{8}}\big{[}&\vert0000000\rangle+\vert1010101\rangle+\vert0110011\rangle+\vert1100110\rangle \\
+ & \vert0001111\rangle+\vert1011010\rangle+\vert0111100\rangle+\vert1101001\rangle\big{]}
\end{array}}
\end{equation}
\begin{equation}
{\setlength\arraycolsep{1pt}
\begin{array}{rl}
\vert1_{L}\rangle=\frac{1}{\sqrt{8}}\big{[}&\vert1111111\rangle+\vert0101010\rangle+\vert1001100\rangle+\vert0011001\rangle \\ + & \vert1110000\rangle+\vert0100101\rangle+\vert1000011\rangle+\vert0010110\rangle\big{]}.
\end{array}}
\end{equation}

\section{Logical state preparation}
\label{LSP}

\begin{figure}[h!]
\begin{center}
$\begin{array}{c} \raisebox{6.5pt}{$\vert0\rangle$} \\ \raisebox{6.5pt}{$\vert0\rangle$} \\ \raisebox{6.5pt}{$\vert0\rangle$} \\ \raisebox{6.5pt}{$\vert0\rangle$} \\ \raisebox{6.5pt}{$\vert0\rangle$} \\ \raisebox{6.5pt}{$\vert0\rangle$} \\ \raisebox{6.5pt}{$\vert0\rangle$} \end{array}$ \raisebox{-65.0pt}{\includegraphics[width=0.36\textwidth]{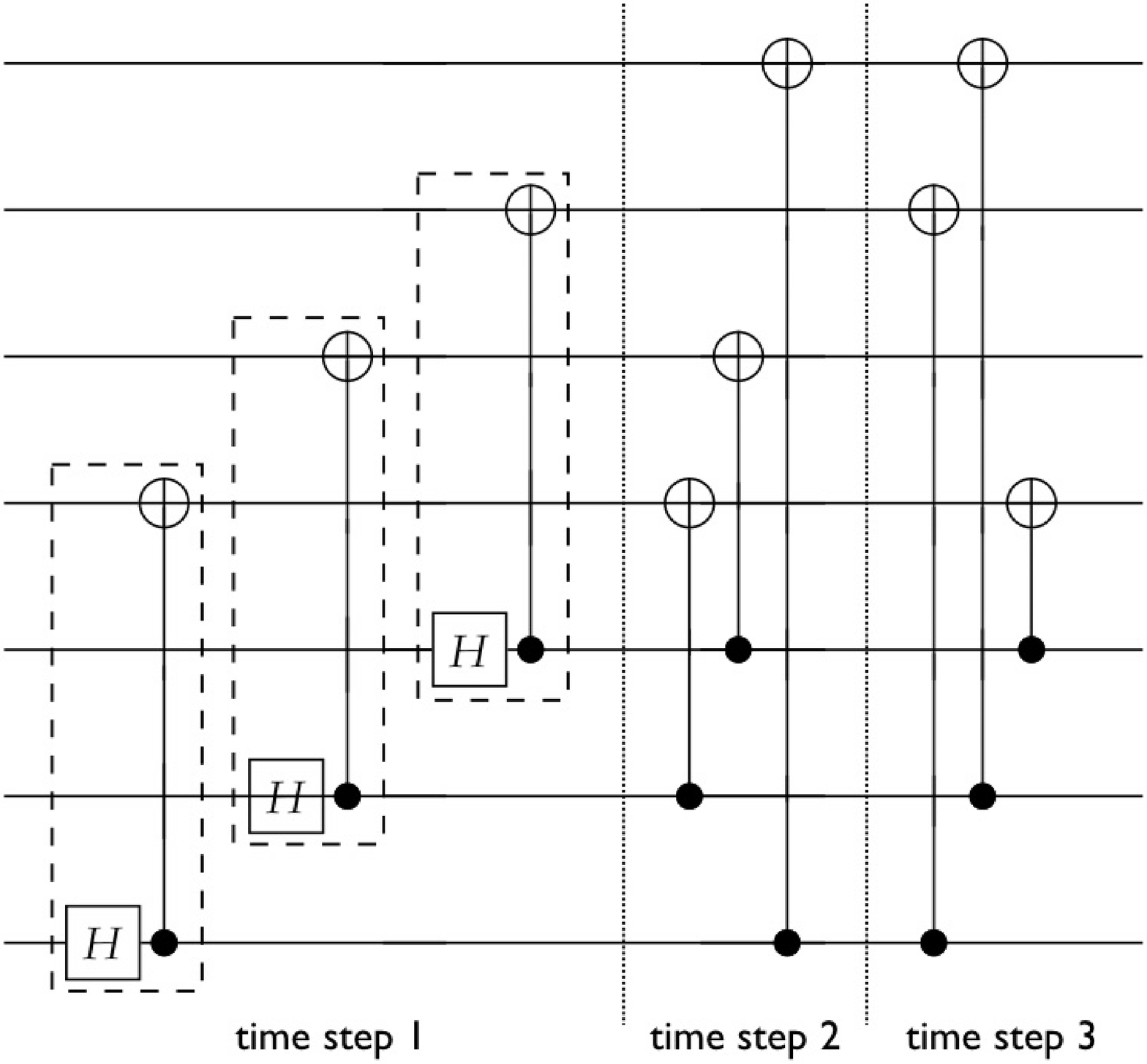}} $\begin{array}{c} \raisebox{5.0pt}{$\vert0_{L}\rangle$} \end{array}$
\end{center}
\vspace*{0pt}
\fcaption{\label{dci}The direct circuit for seven qubit logical-zero preparation, where dashed boxes represent gates which can, in principle, be compounded via canonical decomposition \cite{Sastry1, Cirac2, Makhlin1}.}
\end{figure}

The preparation of logical states is a prerequisite for QEC and forms a basis for the proceeding comparison of non fault-tolerant and fault-tolerant quantum computation. Figure \ref{dci} outlines the direct circuit for the preparation of the Steane seven qubit logical-zero. For brevity, the corresponding fault-tolerant circuit is omitted from this paper, but is detailed elsewhere \cite{Nielsen1}. In the formulation of these circuits it is assumed that non-local interaction between qubits is permitted and that isolated two-qubit interactions can be performed simultaneously. 
\\
The direct implementation comprises a sequence of three time steps, shown explicitly in Figure \ref{dci}, where combined gates are constructed via canonical decomposition \cite{Sastry1, Cirac2, Makhlin1}. In the more complex fault-tolerant circuit logical preparation is achieved by the operator measurement of the stabilizers of the Steane logical codewords \cite{DiVincenzo2, Gottesman1}. However, to maintain fault-tolerance during preparation, each data qubit must interact with a unique ancilla qubit. Therefore, preceding each operator measurement, a separate ancilla block is initialised and verified, whereupon any deviation from the desired ancilla state will ideally result in the ancilla being reset and the initialisation repeated. A number of protocols exist for the preparation of an appropriate ancilla network, including those optimised for accuracy \cite{Reichardt1, Steane5} and those making use of quantum teleportation \cite{Knill2}. To minimise classical memory requirements, Shor's ancilla \cite{Shor3, Preskill2} is chosen, in which a four qubit ancilla is initialised to the maximally entangled state $\frac{1}{\sqrt{2}}\left(\vert0000\rangle+\vert1111\rangle \right)$. A fifth ancilla qubit is used to verify the fidelity of this ancilla state prior to each syndrome measurement. To protect against $Z$ errors propagating to the ancilla block, each stabilizer operator is measured multiple times, after which a majority result is taken to specify the syndrome. The significant resource overhead associated with fault-tolerant computation is quantified in the following table contrasting direct and fault-tolerant logical state preparation.
\begin{figure}[h!]
\begin{center}
\begin{tabular}{lllll}
circuit & number of qubits & circuit depth & circuit area & number of gate operations\\
\hline
direct & 7 & 3 & 21 & 9 \\
fault-tolerant & 12 & $\geq60$ & $\geq720$ & $\geq108$ \\
\hline
\end{tabular}
\vspace*{0pt}
\end{center}
\end{figure}

\section{Classical simulation}
\label{SM}
Classical simulation of the preceding circuits involves storage of the system density matrix, $\rho$, where gate-level manipulation of the system is achieved through the application of appropriate unitary operations. For a two-level, $N$ particle quantum system the representative density matrix is of dimensionality $2^N$, hence a significant memory overhead is associated with even a small number of qubits. Classical simulation is made possible via the implementation of a sparse matrix storage protocol, whereby the density matrix is recast in a tree-like structure with computational memory only being allocated to non-zero matrix elements.  To generate this structure the density matrix is divided successively into quadrants, where each quadrant notionally forms a tree branch.  Branch subdivision will only continue if a quadrant contains any non-zero elements, otherwise the branch is truncated.  Consequently, storage of only the non-zero matrix elements is required, along with their address within the matrix.
\\
\\
Quantum errors are assumed to be stochastic and uncorrelated, such that, at any given time, a discrete error is induced at any location within a circuit with a probability $p$. Errors are applied to qubits involved in gate operations and also to idle qubits, such that errors occur with equal probability on every qubit after every time step within the circuit. Additionally, each error within the set $\{X,Z,XZ\}$ is assumed to occur with an equal probability given by $p/3$. Following the operator-sum formalism \cite{Nielsen1, Krauss1}, the transformation used in the simulation of quantum errors is given by
\begin{equation}
\rho\longrightarrow\rho^{\prime}=(1-p)\rho+\frac{p}{3}\left(X\rho X^{\dagger}+Y\rho Y^{\dagger} + Z\rho Z^{\dagger} \right).
\label{transform}
\end{equation}
Given a knowledge of the correctable error set $\{X,Z,XZ\}$, a set of permissible states is known, each of which can be rectified to yield the exact logical-zero. To determine the fidelity of each simulated circuit its output is measured against this set, which comprises the exact logical-zero, where no error has occurred, plus a further 63 states which differ from this state by either a single $X$ error (7), a single $Z$ error (7), or a combination of a single $X$ error and a single $Z$ error applied to arbitrary qubits (49). Logical preparation is defined to be successful if any one of these 64 states is generated, a definition which implicitly assumes that computation can continue following the appropriate correction circuit. Labeling as $\rho_{i}$ the density matrices representing each of these correctable states, for a given output, $\rho_{s}$, the fidelity of the circuit is defined by
\begin{equation}
\vert\displaystyle\sum_{i} \rm{tr}\big(\rho_{s}\rho_{i}\big)\vert^{2}.
\label{fidelity}
\end{equation}
Finally, provisions are made when simulating the fault-tolerant circuit to account for the possibility that ancilla preparation may be repeated and that syndrome measurements may occur two or three times. This dynamic circuit structure is modeled by a list of classical measurement scenarios, each described by a binary string. At each measurement gate in the circuit, a digit in the string is used to force the measurement outcome. By tracking the probability of obtaining the measured outcome at each gate, the overall probability of a particular scenario eventuating can be determined. Scenarios of decreasing likelihood are simulated in this way, with the fidelity of each scenario weighted by its corresponding probability to form an overall circuit fidelity.

\section{Results}
\label{A}

Simulations were initially undertaken to determine the susceptibility of the direct implementation to specific single qubit errors. Consistent with the error model, single qubit errors were manually applied after each of the three time steps. As expected, errors were observed to propagate within the circuit to afflict multiple qubits, though these accumulated errors did not invariably lead to circuit failure. Rather, only six errors were observed to generate more than one effective error at the output. For every other error, the cumulative effect of any propagated errors was equivalent to that of only a single error, though not necessarily on the same qubit as was the initial error. This reduction in the effective area of the circuit was attributed to symmetries inherent the code structure. For low $p$, where the probability of two errors occurring is negligible, the circuit fidelity was analytically calculated as $1-4p+O(p^2)$. This result was in complete agreement with simulated results and provided a non-trivial test of the method of simulation.
\\
\\
The fidelities of the both the direct and fault-tolerant circuits for logical-zero preparation were evaluated for varying values of the physical error rate $p$. Results confirmed the relative stability of the direct circuit in the high $p$ regime, as is illustrated in Figure 2(a). In this region, the significant complexity introduced in the fault-tolerant circuit increases dramatically the probability that multiple errors will occur. For lower values of $p$, it was expected that the fault-tolerant circuit would fail with probability $cp^2$, where $c$ is the number of pairs of locations at which two errors can occur. By convention, a naive estimate of $c$ can be made by considering the circuit area $A$, such that
\begin{figure}
\begin{center}
\subfigure[Fidelity of the direct and fault-tolerant circuits for logical-zero preparation in the presence of quantum errors. Results shown span the high $p$ regime, in which the direct circuit is relatively robust.]{\includegraphics[width=0.45\textwidth]{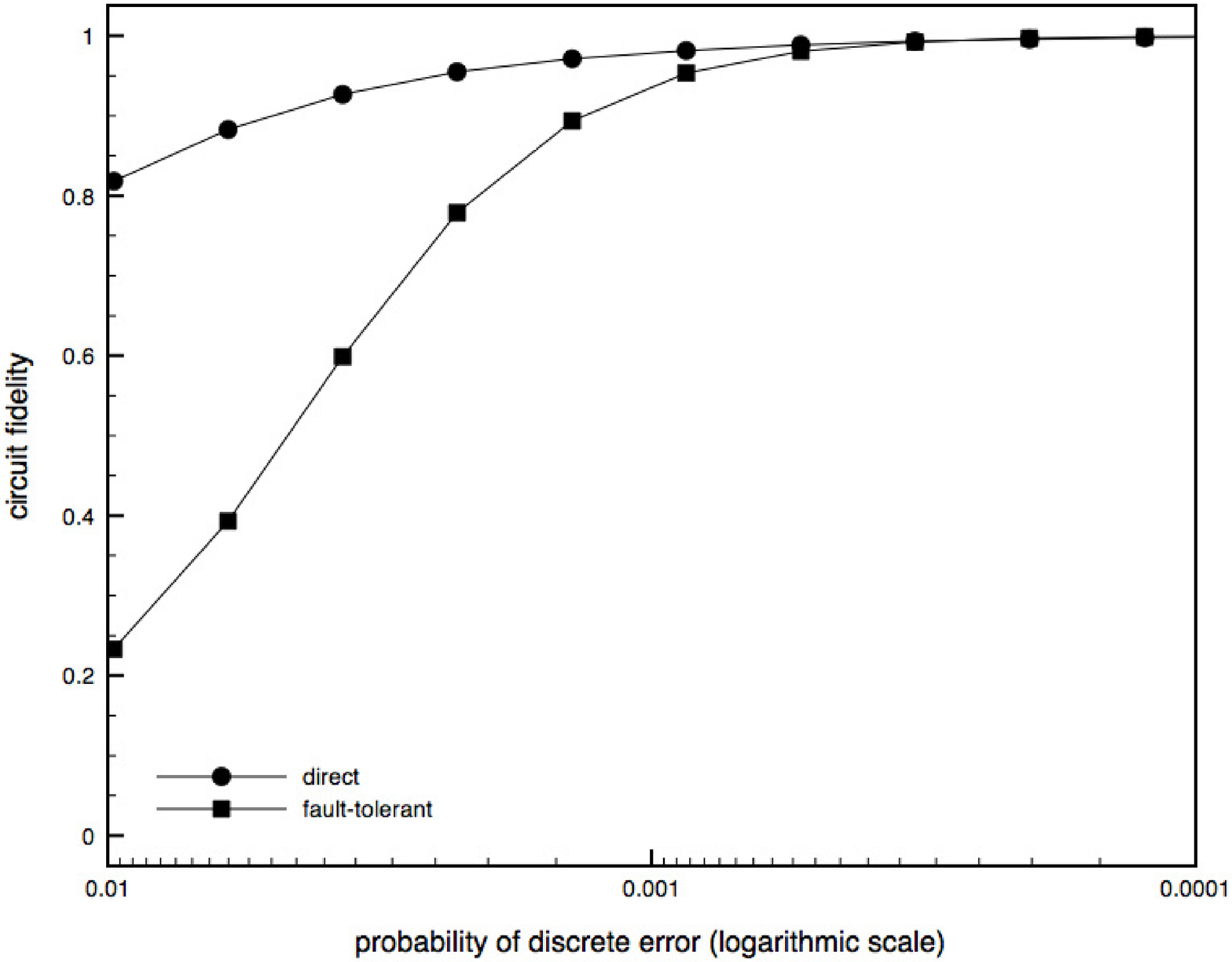}} \hspace{1.0cm}
\subfigure[The crossover error rate at which fault-tolerant and direct preparation become effectively equivalent. For physical error rates below $5.3\times10^{-5}$ the fault-tolerant circuit achieves logical state preparation with a greater reliability than the equivalent direct implementation]{\includegraphics[width=0.45\textwidth]{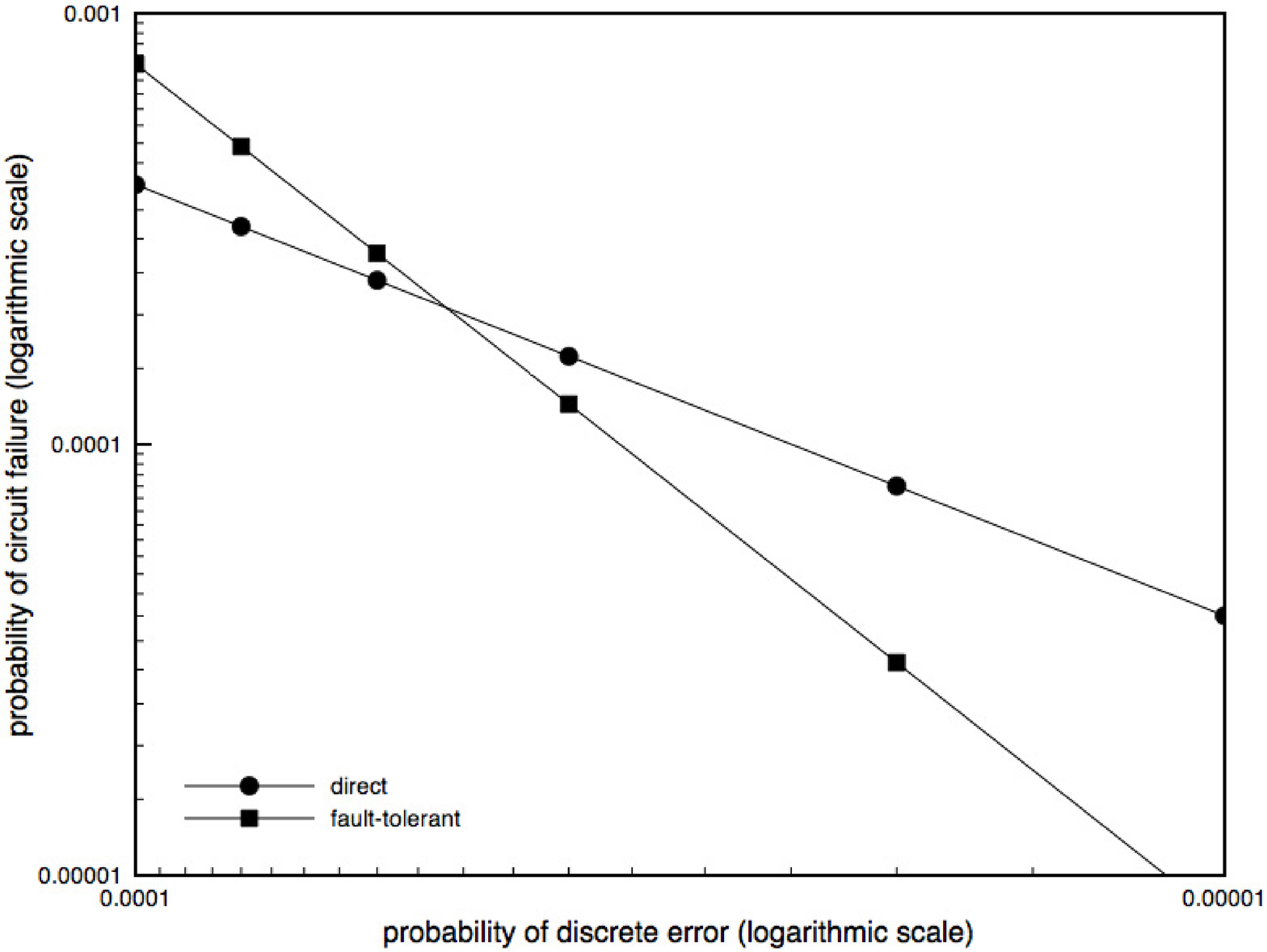}}
\fcaption{\label{dual}}
\end{center}
\end{figure}
\begin{equation}
c\approx \left( \begin{array}{c} A \\ 2 \end{array} \right) \geq \left( \begin{array}{c} 720 \\ 2 \end{array} \right) = 2.6\times10^5.
\end{equation}
The fault-tolerant circuit was found to be more robust than predicted by this simple estimation. Based upon the simulated data, $c$ was determined to equal approximately $7.7\times10^4$, implying that the fault-tolerant circuit does not fail upon the application of every two error combination. While the observed reduction in the effective area of the fault-tolerant circuit is partly analogous in both cause and effect to that observed of the direct circuit, this reduction can also be attributed to the fault-tolerant structure of both ancilla preparation and syndrome extraction. For example, as certain errors during preparation can be detected, these errors will not contribute to the set of errors which cause circuit failure. This general result further serves to illustrate a significant advantage of gate-level simulation, as it facilitates a direct observation of the response of a complex circuit to quantum errors. Moreover, the quantitative analysis of error combinations has been tied previously to a rigorous determination of the fault-tolerant threshold \cite{Aliferis1}.
\\
\\
The crossover physical error rate at which the direct and fault-tolerant circuits operated with equal fidelity was observed at $p_{cr}\approx5.3\times10^{-5}$. Figure 2(b) illustrates the crossover and highlights the behaviour of each circuit in the low $p$ regime. As expected, in this region the direct circuit fails with probability $O(p)$, while the corresponding fault-tolerant circuit fails with probability $O(p^2)$. Assuming that a quantum computer will operate with an error rate of approximately $p_{cr}$ then implies that fault-tolerant preparation may enable logical encoding of data with a higher fidelity than direct preparation. In contrast, the compact nature of the direct circuit will allow more logical states to be prepared for use as ancillas in a given cycle of QEC. Recent research has demonstrated that non-fault tolerant ancilla preparation, coupled either with a verification procedure \cite{Svore1} or with a decoding procedure \cite{DiVincenzo3}, is important in achieving an attractive threshold when considering an ancilla network with logical states as the primary resource.

\section{Conclusion}
\label{C}

A crossover physical error rate was observed at approximately $5.3\times10^{-5}$, below which the fault-tolerant implementation becomes the preferred method of state preparation. The complexity associated with the fault-tolerant circuit clearly provides the major impediment to its stability under quantum errors. Though fault-tolerant preparation is, at present, unable to operate effectively in the region of realistic physical error rates, fault-tolerant computation is necessary to achieving the increased fidelity afforded by concatenated QEC. Thus, while higher order QEC codes may prove to be more robust, more rapid and efficient fault-tolerant circuits must be designed that take advantage of non fault-tolerant components where appropriate such that maximum stability is achieved. This process should be aided by the simulation of other direct and fault-tolerant circuits, including correction and transversal gate operation.
\\
\\
In undertaking further simulation, greater consideration must be given to the environment in which quantum computation will inevitably take place. To this effect, recent threshold analyses have been undertaken in the context of a specific spatial arrangement of qubits \cite{Szkopek1, Svore1}. In addition, while the model described in Section \ref{SM} is common within QEC analyses, and in spite of previous research investigating the effects of local non-Markovian noise on computation \cite{Terhal1, Aliferis1}, a more thorough error model should more accurately reflect the physical properties of a specific computing architecture. For example, a given architecture may be more vulnerable to dephasing errors, or may experience high rates of qubit loss. Errors may also be attributed to inaccurate gate implementation, in which the assumption of non-correlated error effects will be invalidated, and errors associated with qubit transport must accompany any analysis of concatenated architectures, which are vital to any practical realisation of the threshold theorem. Such spatial and physical considerations not only highlight the challenge of accurately and realistically simulating quantum circuits, but demonstrate explicitly the need for circuits conceived in the context of specific physical architectures.

\section{Acknowledgements}

This work was supported by the Australian Research Council, the Australian Government, and by the US National Security Agency (NSA), Advanced Research and Development Activity (ARDA), and the Army Research Office (ARO) under contract number W911NF-04-1-0290.

\nonumsection{References}
\bibliographystyle{unsrt}
\bibliography{bib2.bib}



\end{document}